# Engineering spin wave spectra in thick $Ni_{80}Fe_{20}$ rings by using competition between exchange and dipolar fields

X. Zhou,[1] E.V.Tartakovskaya,[2,3,*] G.N.Kakazei,[4] and A. O. Adeyeye[1,5,†]

[1]Department of Electrical and Computer Engineering, National University of Singapore, 4 Engineering Drive 3, Singapore 117576, Singapore

[2]Faculty of Physics, Adam Mickiewicz University, Poznan, Uniwersytetu Poznanskiego 2, 61-614 Poznan, Poland

[3]Institute of Magnetism, National Academy of Sciences of Ukraine, 36b Vernadskogo Blvd, 03142 Kiev, Ukraine

[4]Institute of Physics for Advanced Materials, Nanotechnology and Photonics (IFIMUP)/ Departamento de Fisica e Astronomia, Universidade do Porto, 4169-007, Porto, Portugal

[5]Department of Physics, Durham University, South Rd, Durham, DH1 3LE, UK

**Abstract.** Control of the spin wave dynamics in nanomagnetic elements is very important for the realization of a broad range of novel magnonic devices. Here we study experimentally the spin wave resonance in thick ferromagnetic rings (100 nm) using perpendicular ferromagnetic resonance spectroscopy. Different from what was observed for the continuous film of the same thickness, or from ring with similar lateral dimensions but with lower thicknesses, the spectra of thick patterned rings show a non-monotonic dependence of the mode intensity on the resonance field for a fixed frequency. To explain this effect, the theoretical approach by considering the dependence of the mode profiles on both the radial and axial coordinates was developed. It was demonstrated that such unusual behavior is a result of the competition between exchange and dipolar fields acting at the spin excitations in the structure under study. The calculations are in a good agreement with the experimental results.

---

[*]olena.tartakivska@gmail.com

[†]adekunle.o.adeyeye@durham.ac.uk

# I. Introduction

Standing spin waves (SW) are being studied in magnetic materials since the late fifties of 20$^{th}$ century, when Charles Kittel theoretically predicted their existence in perpendicularly magnetized ferromagnetic films [1] and almost immediately they were observed experimentally [2]. With the development of patterning techniques, particularly electron-beam lithography, standing SW modes were observed in micron-sized ferromagnetic elements due to their restricted lateral dimensions [3-7]. Having in mind that magnetic nanoelements are being used in many real world applications – magnetic recording read-write heads [8], magnetic random access memories [9], spin-torque oscillators [10], and microwave to spin-wave transducers [11] just to name a few – it is crucial to know and understand their SW spectra. So far the main attention was given to thin infinite stripes [3,4] and circular disks [5,6,12-15], mainly due to the possibility to describe their SW eigenfunctions analytically. The approximate solutions for the eigenfunctions of the stripes had sinusoidal form, and for circular dots the spin wave profiles were satisfactorily described by the zeroth-order Bessel functions of the first kind.

The discovery of new fascinating properties of ferromagnetic rings (vortex and onion magnetic states below saturation, specific features of the switching behavior and peculiarities of magnetic dynamics) [16-21], as well, as the possibility to control vortex chirality in rings [22,23], make them a promising candidate for usage in magnetoresistive random access memory (MRAM) devices [24,25] and logic circuits [26]. Very recently nanovolcanoes - nanodisks overlaid by nanorings – were proposed as purpose-engineered 3D architectures for nanomagnonics [27]. Obvious advantage of magnetic rings when compared with dots is the option to vary the internal diameter, which provides additional possibility to control their magnetic properties, including the SW spectra. This explains the extensive investigation of magnetic rings during the last twenty years. However, in previous studies of magnetic dynamics, the main attention was paid to thin rings (with thickness below 50 nm) [28-32].

Ferromagnetic rings have axial symmetry similar to dots, but the topological non-equivalence of the shape of dots and rings leads to specific differences of corresponding SW spectra even in the case of thin nanoparticles [32]. The additional effects arising due to varying of the thickness of nanoparticles can be explained if we compare dimensions of SW profiles in films and nanoparticles with corresponding material's parameters, i.e., the characteristic lengths of exchange and dipolar interactions. The main difference between the standing SWs in continuous films (where wave vectors K are oriented along the normal to the film plane) and in micron-sized patterned elements (where K are oriented along the film plane) is the distance between SW maximums – in the former case (thickness, 50~100 nm) it is comparable with material exchange length and in the latter (lateral dimensions, 500 ~ 2000 nm) it is noticeably bigger. However, with further advances in lithography, the difference between the element thickness and lateral dimensions became smaller, and therefore both the exchange and dipolar interactions are playing equally important roles. It is expected that the intensity of the spin-wave modes decreases with the increase of mode number due to the diminishing of net magnetization. Recently it was demonstrated that this simple rule does not apply for the vortex gyrotropic modes in thick circular magnetic dots [33,34] due to the complex thickness profiles of phases of different SW modes. However, for the SWs in magnetically saturated samples so far this was always the case.

In this work we prepared ferromagnetic rings array by combining lithography (ring width 400 nm, dipolar limit) with a large thickness (100 nm, exchange limit). Opposite to the majority of previous investigations, we consider the saturated state, when the external magnetic field is applied perpendicularly to the rings' plane (perpendicular geometry). The observed SW modes are inhomogeneous along the sample thickness and their intensities are not reciprocally proportional to mode number, as in the typical SW resonance, that was attributed to the competition between exchange and dipolar fields. A simple analytical approach presented here fits quantitatively the experimental data.

## II. Sample fabrication and measurements details

The periodic array of $Ni_{80}Fe_{20}$ rings was fabricated on a silicon substrate using deep ultraviolet lithography at 248 nm exposure wavelength followed by electron beam evaporation and ultrasonic assisted lift off process in resist thinner. An adhesion layer of 5 nm Cr followed by 100 nm thick $Ni_{80}Fe_{20}$ film were deposited at a constant rate of 0.4 Å/s with a base pressure of $4 \times 10^{-8}$ Torr. Details of the fabrication process are described elsewhere [35]. The lateral dimensions of the rings under study are: outer radii $R = 1500$ nm and inner radii $r = 1100$ nm. The inter-ring center-to-center distance is equal to $4R = 6000$ nm, which is sufficient to exclude the influence of inter-ring magnetostatic interactions on the resonance peak positions. Scanning electron microscope (SEM) was used to verify the dimensions of the rings and the sharpness of ring edges (Fig. 1). The bottom left inset is the corresponding SEM image of the individual ring. A 100 nm thick $Ni_{80}Fe_{20}$ reference film was also prepared in the same fabrication cycle. Its saturation magnetization $M$ was found to be 775 emu/cm$^3$ from broadband ferromagnetic resonance measurements.

The microwave absorption of the samples was studied using broadband perpendicular ferromagnetic resonance spectroscopy (FMR) at room temperature. The special attention was given to the precise alignment of the external magnetic bias field $H_{op}$ along the normal to the sample plane. Microwave signal was generated by a continuous wave microwave generator at a specific frequency in the range from 6 GHz to 13 GHz. The samples were placed on top of a 50 Ω microstrip line with the ring arrays facing the line. The width of the microstrip line was 3 mm, the size of the sample – 5×5 mm, and the intensity of microwave signal on the sample was estimated to ne 2 dBm. $H_{op}$ was additionally modulated with an ac field ($H_{ac}$) of 20 Oe, generated by the pair Helmholtz coils placed around dc magnet poles. The output dc signal of the interferometric detector was fed into a digital lock-in amplifier which was locked to the $H_{ac}$ modulation signal. The FMR signal detected in this way represents the first derivative of the field sweeping absorption curve at a selected frequency. For each frequency $H_{op}$ was swept from 18 kOe to 0. A sketch of the field geometry is shown as the top right inset in Fig. 1.

### III. Results

The microwave absorption spectrum for a 100 nm thick $Ni_{80}Fe_{20}$ continuous film at $f = 11$ GHz is shown in Fig. 2(a). Three distinct resonance modes are clearly observed ($i = 1$: $H = 13.51$ kOe; $i = 2$: $H = 12.73$ kOe; $i = 3$: $H = 11.78$ kOe). The mode intensity gradually decreases with the increasing mode number. This behavior is attributed to the quantization of spin wave modes along the film's thickness. The spectrum for 100 nm thick $Ni_{80}Fe_{20}$ rings at the same frequency is shown in Fig. 2(b). In this case, four resonance modes are observed. Different from what was found for the continuous film, for thick rings the first mode ($i = 1$) observed at $H = 11.07$ kOe shows a smaller absorption amplitude than the second mode ($i = 2$) appearing at $H = 10.14$ kOe. Besides the two modes and the next intense mode ($i = 4$, $H = 8.76$ kOe), we also observed an additional mode of small intensity to the right of the second mode ($i = 3$, $H = 9.80$ kOe).

Shown in Fig. 3 is the experimental dispersion relation (6 ~ 13 GHz) for the 100 nm thick continuous film fitted by the well-known Kittel formula,

$$f_K = \frac{\gamma}{2\pi}\left[H_{op} - 4\pi M + \frac{2A}{M}K^2\right]. \qquad (1)$$

Here $A$ is the exchange stiffness constant, and $K$ is the out-of-plane wave vector of the spin wave excitations. Standard $Ni_{80}Fe_{20}$ values were selected for the parameters $A$ and $\gamma$, $A = 1.3\times10^{-6}$ erg/cm, $\gamma/2\pi = 2.93$ MHz/Oe, while the quantized values of $K$ were extracted from the experimental results to describe the field gaps between the three observed modes, $K_1 = 0.47\pi/a$, $K_2 = 1.4\pi/a$, $K_3 = 2.3\pi/a$, where $a = 100$ nm is the film thickness. Such values of $K$ are not standard for spin wave resonance with "pinning" conditions at the surfaces $z = \pm a/2$ (when quantized values of $K$ for detectable modes should be $K_1 = \pi/a$, $K_2 = 3\pi/a$, $K_3 = 5\pi/a$), or for spin wave resonance with "free" boundary conditions (when only one homogeneous "Kittel" mode with $K_1 = 0$ could be observed in experiment). Therefore, mixed boundary conditions take place in this case. If $K_2 - K_1$ and $K_3 - K_2$ are close to $2\pi$, the equal pinning forces act at the both surfaces of the film, and the boundary conditions for spin wave excitations can be described by only one pinning parameter. However in our case such values are much smaller (actually $K_2 - K_1$ and $K_3 - K_2$ are close to $\pi/a$). This experimental observation gives the evidence of different pinning at the upper and lower film's surfaces, which strengths can be described by two different pinning parameters, $\sigma_1 = 0.1$ and $\sigma_2 = 10$ (see Appendix, Ref. 36 and references therein). Such difference is most probably caused by the fact that NiFe film was deposited on top of Cr underlayer, however no protective layer was deposited on top of NiFe film. As a result, the bottom surface of NiFe film is free from oxide layer (remnant oxygen on the substrate was absorbed by Cr film), therefore pinning is almost absent. On the contrary, the top surface of NiFe film was exposed to the air during chamber venting, which led to its oxidation, and, subsequently, to the noticeable pinning parameter. The profile of the spin wave excitation along the film's thickness can be chosen in the next form, $A_n \cos(K_n z) + B_n \sin(K_n z)$, where the amplitudes $A_n$ and $B_n$ depends on the wave vector of the spin wave. In the Appendix we give detailed calculations of the relation $\frac{A_n}{B_n}$ for three experimentally observed branches of spin wave excitations. The values $(A_1/B_1)^2 = 1.54$, $(A_2/B_2)^2 = 0.4$, $(A_3/B_3)^2 = 3.96$ (see formulae (12A), (13A)), define the corresponding intensities, $I(K_n) = \dfrac{\left(\dfrac{A_n}{B_n} \int\limits_{-a/2}^{a/2} \cos(K_n z) dz\right)^2}{a \int\limits_{-a/2}^{a/2} \left(\dfrac{A_n^2}{B_n^2} \cos(K_n z)^2 + \sin(K_n z)^2\right) dz}$. The obtained relations between intensities $I(K_1) : I(K_2) : I(K_3) = 1 : 0.11 : 0.03$ are in agreement with experimental values.

While the analytical theory of SW in thin circular magnetic dots was developed more than ten years ago for both saturated [5] and non-saturated [37] states, this problem turned out to be much more complicated in rings. Just recently we presented the first analytical calculations of SW spectra in thin (30 nm) ferromagnetic rings [32], where the spin wave profile was considered to be a function of only the in-plane (radial and azimuthal) coordinates. This is the standard approach to the analysis of spin excitations in planar nanostructures (see [5,38-43] and references therein) that gives the monotonic decrease of the mode intensity with mode numbers. In the case of thick rings this consideration is not valid since it is necessary to take into account the inhomogeneity of the spin wave profile along two spatial variables, radial coordinate $\rho$ and axial coordinate $z$. We assume, that in the cylindrical geometry the shape of eigenmodes of thick rings can be described by products of $\rho$-dependent and $z$-dependent components,

$$\mu_{m,n}(\rho, z) = \mu_m(\rho) \cdot \left[ A_n \cos(K_n z) + B_n \sin(K_n z) \right] \qquad (2)$$

where $\mu_m(\rho)$ can be performed as a linear combination of zeroth-order Bessel functions of the first and the second kind,

$$\mu_m(\rho) = \left[ J_0(k_m \rho) + C_m Y_0(k_m \rho) \right] \qquad (3)$$

We also assume, that the functions $\mu_m(\rho)$ satisfy pinning conditions at the ring's edges, $\mu_m(\rho = R) = \mu_m(\rho = r) = 0$. Full pinning is actually an approximate assumption for the nanostructures whose thickness is not extremely small in comparing with the lateral dimension. However our preliminary estimations proved that in the present case this approach is reliable, as additional corrections of boundary conditions on the sides of the ring does not change the result significantly. Application of the pinning conditions leads to the system of two equations, from which the quantized values of the in-plane wave vector $k_m = \frac{\beta_m}{R}$, and the corresponding coefficients $C_m = \frac{J_0(\beta_m)}{Y_0(\beta_m)}$ can be found, where $\beta_m$ is a root of the relation

$$J_0(\beta_m)Y_0\left(\beta_m \frac{r}{R}\right) - Y_0(\beta_m)J_0\left(\beta_m \frac{r}{R}\right) = 0 \tag{4}$$

So, due to the system's confinement in radial and axial directions, every spin wave mode profile corresponds to the set of two quantized values of wave vectors ($k_m$, $K_n$). Further, we have used the set of two numbers ($m,n$), to identify the mode profile $\mu_{m,n}(\rho, z)$, the corresponding frequency $f_{mn}$ and the effective matrix element of the inhomogeneous demagnetizing field $N_{mn}$.

The effective demagnetizing factor of the ring along the normal direction is also a function of radial and axial coordinates,

$$N(\rho, z) = \frac{1}{4\pi} \frac{\partial}{\partial z} \int_V d\mathbf{r}' \frac{\partial}{\partial z'} \frac{1}{|\mathbf{r} - \mathbf{r}'|} \tag{5}$$

Applying the formula

$$\frac{1}{|\mathbf{r} - \mathbf{r}'|} = \sum_{m=-\infty}^{\infty} \int_0^{\infty} dt\, e^{im(\varphi - \varphi')} J_m(t\rho) J_m(t\rho') e^{-t|z-z'|} \tag{6}$$

we receive for $N(\rho, z)$ of the ring with the inner radius $r$ and the outer radius $R$ the next expression,

$$N(\rho, z) = \int_0^{\infty} dt\, (RJ_1(Rt) - rJ_1(rt))\, J_0(t\rho) e^{-ta/2} \cosh(tz) \tag{7}$$

To explain the experimentally observed FMR spectra of rings (Fig.2(b)), we use Herring-Kittel spin-wave dispersion relation [44], appropriately modified for the case of the ring-shape magnet with the inhomogeneous spin wave profile along $z$,

$$f_{mn}^2 = \left(\frac{\gamma}{2\pi}\right)^2 \left[H - 4\pi M N_{mn} + \frac{2A}{M}(k_m^2 + K_n^2)\right] \left[H - 4\pi M N_{mn} + \frac{2A}{M}(k_m^2 + K_n^2) + 4\pi M F(k_m, K_n)\right] \tag{8}$$

Here, $N_{mn}$ is the effective matrix element of the inhomogeneous demagnetizing field for different standing spin wave modes [45],

$$N_{mn} = \frac{1}{\int_{-a/2}^{a/2} dz \int_r^R d\rho \rho (\mu_{m,n}(\rho,z))^2} \int_{-a/2}^{a/2} dz \int_r^R d\rho \rho (\mu_{m,n}(\rho,z))^2 N(\rho,z) \qquad (9)$$

which can be calculated by employing the formulae (2), (3), (7). The matrix element of dipole-dipole interaction for spin waves with inhomogeneous profile along rings' thickness takes the form

$$F(k_m, K_n) \frac{\frac{A_n^2}{B_n^2} F_1(k_m, K_n) + F_2(k_m, K_n)}{\int_{-a/2}^{a/2} \left( \frac{A_n^2}{B_n^2} \cos(K_n z)^2 + \sin(K_n z)^2 \right) \frac{dz}{a}}, \qquad 10)$$

where

$$F_1(k,K) = \frac{k^2}{2(k^2+K^2)} \cdot \left( \left(1 + \frac{\sin(Ka)}{Ka}\right) - \frac{1}{ka} \left[ \begin{array}{l} \left(1 - e^{-ka}\cos(Ka)\right)\left(\frac{(k^2-K^2)}{(k^2+K^2)} + \cos(Ka)\right) + \\ + e^{-ka}\sin(Ka)\left(\frac{2Kk}{(k^2+K^2)} - \sin(Ka)\right) \end{array} \right] \right) \qquad (10a)$$

$$F_2(k,K) = \frac{k^2}{2(k^2+K^2)} \cdot \left( \left(1 - \frac{\sin(Ka)}{Ka}\right) - \frac{1}{ka} \left[ \begin{array}{l} \left(e^{-ka}\cos(Ka)-1\right)\left(\cos(Ka) - \frac{(k^2-K^2)}{(k^2+K^2)}\right) + \\ + e^{-ka}\sin(Ka)\left(\frac{2Kk}{(k^2+K^2)} + \sin(Ka)\right) \end{array} \right] \right) \qquad (10b)$$

In the case $K = 0$ (i.e., when the mode's profile is homogeneous along z), $F(k,0) = 1 - \frac{(1-e^{-ka})}{ka}$, i.e. coincides with the matrix element of dipole-dipole interaction for perpendicularly magnetized thin films and for planar elements like circular dots [5,33,46].

The first five roots of (4) can be calculated numerically. In the case of $r = 1100$ nm and $R = 1500$ nm, we have $\beta_1 = 11.77$, $\beta_2 = 23.554$, $\beta_3 = 35.337$, $\beta_4 = 47.12$, $\beta_5 = 58.902$. The corresponding profiles of the radial component, $\mu_m(\rho)$, calculated by formula (3), are presented in Fig. 4. Profiles with even indices (corresponding to in-plane wave vectors $k_2 = \frac{\beta_2}{R}$ and $k_4 = \frac{\beta_4}{R}$), which are antisymmetric functions of $\rho$, cannot be detected by experimental techniques, while modes with odd indices ($m = 1, 3, 5$) can be observed experimentally. Althoughwe demonstrate this numerical result for given values of $r$ and $R$, such rule is generally valid for spin wave excitations in rings in the perpendicular geometry.

## IV. Discussion

It is natural to assume that the boundary conditions for rings at the surfaces $z = \pm a/2$ are similar to the boundary conditions for the reference film. Hence, the corresponding values of out-of-plane wave vectors $K_i$ are close to the set that we defined before for the 100 nm thick continuous film, i.e., $K_1 = 0.47\pi/a$ and $K_2 = 1.4\pi/a$.

Presented in Fig. 5 is the comparison between the calculation results of $f_{mn}$ with indices ($m, n$) = (1, 1), (1, 2), (3, 1), (3, 2) and the experimental data. Theoretically calculated frequencies of standing spin waves are in a good agreement with the experimentally observed values for all the four modes. However, the most interesting feature of these calculations is that the mode with indices ($m, n$) = (1, 1) is excited by a lower resonance field than the mode with indices ($m, n$) = (1, 2), and, similarly, the resonance field of the mode ($m, n$) = (3, 1) is lower than of the mode ($m, n$) = (3, 2). As the intensity of the FMR signal is proportional to the integral by the spatial coordinates from the mode's profile, this means that the mode $i = 1$ {($m, n$) = (1, 2)} has a smaller intensity than the mode with index $i = 2$ {($m, n$) = (1, 1)}, but a higher resonance field, i.e. in comparing with data of the film, rings have the inverse dependence for field vs mode's number $i$. This theoretical conclusion is in a qualitative agreement with the unusual dependence of the mode intensity as aforementioned (Fig.2(b)).

Next we will show that the theoretical approach, developed in the paper allows us to explain the physical origin of such effect. As the demagnetizing field in infinite films $-4\pi M$ does not depend on $K_n$ while $F(k = 0, K) = 0$ for any $K$ (formula (10a)), the field gaps between different modes in the films are defined only by the exchange interactions, i.e. the term $\frac{2A}{M}K_n^2$. Evidently, this dependence provides monotonic decrease of the resonance field with the increase in $K_n$. As opposed to the film, the demagnetizing factors $N_{mn}$ and the dipole-dipole matrix elements $F(k_m, K_n)$ in rings are dependent on the mode's shape. Particularly the mode's profile along the film thickness and the corresponding wave vector $K_n$ play the most important role. In Fig. 6 we present the dependence of components of the dipole-dipole field $4\pi F_1(k_m, K)$ and $4\pi F_2(k_m, K)$ as a function of $K$ with fixed $k_1 = \beta_1/R$ and $k_3 = \beta_3/R$ (calculations by the formulae (10a), (10b)). Functions $4\pi F_1(k_m, K)$ decrease significantly with the increase of $K$, while functions $4\pi F_2(k_m, K)$ have non-monotonic behavior, being much smaller than corresponding $4\pi F_1(k_m, K)$. As a result, the matrix element of dipole-dipole interaction $F(k_m, K_n)$, (formula (10)) significantly decreases with the increase of $K$. The demagnetizing factors, (formula (9)) slightly increase with the increase of $K$ (our calculations give the results $4\pi N_{11} = 9.8$, $4\pi N_{12} = 10.41$, $4\pi N_{31} = 9.4$, $4\pi N_{32} = 10.1$). This means that if the exchange term $\frac{2A}{M}(k_m^2 + K_n^2)$ increases the frequencies $f_{mn}$ with growing indices $n$, the terms of dipolar origin, vice versa, reduce the corresponding values of the frequencies. In the case of ring's geometry dipolar terms "win this competition", and modes with larger indices $n$ have smaller frequency at the same excitation field. However, the modes with larger indices $n$ have smaller intensity. This leads to the observed unusual dependence of the FMR signal's intensity from the number of the mode. Based on the theoretical approach described above, we have calculated the intensities of four detectable modes of rings using the formula

$$I_{mn} = \frac{\left(\int_{-a/2}^{a/2} dz \int_r^R d\rho\, \rho\, \mu_{m,n}(\rho, z)\right)^2}{a\left(\int_r^R d\rho\, \rho\right) \cdot \int_{-a/2}^{a/2} dz \int_r^R d\rho\, \rho\, (\mu_{m,n}(\rho, z))^2} \qquad (11)$$

The results are presented in Fig.7 together with the experimental curve of the absorption derivative at $f = 11$ GHz for comparison. As one may see, there is a fair agreement between the theory and experiment.

### V.     Conclusions

Using the broadband ferromagnetic resonance technique, we observed the important peculiarity of standing spin wave modes in isolated 100 nm thick $Ni_{80}Fe_{20}$ rings in the perpendicular geometry in comparing with the reference film, particularly the non-monotonic dependence of the absorption amplitudes from the resonance magnetic field.

The multi-resonance spectra of the 100nm reference film demonstrate that modes with different inhomogeneous profiles along the film thickness are excited. It is natural to assume that the modes with inhomogeneous profiles along the thickness of the ring are detected as well. To describe such kind of excitations, we calculated spin waves dispersion in rings, taking into account the dependence of mode profile from two spatial variables, radial coordinate $\rho$ and axial coordinate $z$. This allowed us to achieve a quantitative agreement between experimental data and calculations of spin wave dispersion. Developed theoretical approach explains the physical reason of the observed effect as a competition between exchange and dipolar fields acting at the spin excitations in confined planar particles.

### Acknowledgements

The research leading to these results has received funding from the Norwegian Financial Mechanism 2014-2021 project no UMO-2020/37/K/ST3/02450. This work was supported by the National Research Foundation, Prime Minister's Office, Singapore under its Competitive Research Programme (CRP Award No. NRFCRP 10-2012-03), the European Union's Horizon 2020 research and innovation program under the Marie Skłodowska-Curie grant No. 644348. G.N.K. acknowledges Network of Extreme Conditions Laboratories - NECL and Portuguese Foundation of Science and Technology (FCT) support through the projects NORTE-01-0145-FEDER-022096 and POCI-0145-FEDER-030085 (NOVAMAG). A.O.A. would like to acknowledge the funding from the Royal Society and Wolfson Foundation.

**Appendix. Calculations of the relation $\frac{A_n}{B_n}$ for different experimentally observed branches of spin wave excitations**

Mixed boundary conditions with the same pinning at the top and at the bottom surface of the film can be described by the equations

$$\begin{cases} \dfrac{\partial m}{\partial z} - \dfrac{\sigma}{a} m = 0 \big|_{z=-a/2} \\ \dfrac{\partial m}{\partial z} + \dfrac{\sigma}{a} m = 0 \big|_{z=a/2} \end{cases}, \quad (A1)$$

where $\sigma$ is a pinning parameter, spin wave excitations are chosen as a superposition of cosinusoidal and sinusoidal functions:

$$\begin{aligned} m &= A_K \cos(Kz) + B_K \sin(Kz) \\ \frac{\partial m}{\partial z} &= -A_K \cdot K \sin(Kz) + B_K \cdot K \cos(Kz) \end{aligned} \quad (A2)$$

Inserting (A2) into (A1) we receive the next system of homogeneous linear equations with unknown coefficients $A_K$ and $B_K$:

$$\begin{cases} A_K K \sin(Ka/2) + B_K K \cos(Ka/2) - \dfrac{\sigma}{a}\left(A_K \cos(Ka/2) - B_K \sin(Ka/2)\right) = 0 \\ -A_K K \sin(Ka/2) + B_K K \cos(Ka/2) + \dfrac{\sigma}{a}\left(A_K \cos(Ka/2) + B_K \sin(Ka/2)\right) = 0 \end{cases} \quad (A3)$$

From (A3) we find the transcendental equation [36]

$$tg(Ka) = \frac{2Ka\sigma}{\left[K^2a^2 - \sigma^2\right]}, \tag{A4}$$

while

$$\cos(Ka)^2 = \frac{\left[K^2a^2 - \sigma^2\right]^2}{\left[K^2a^2 + \sigma^2\right]^2}, \tag{A5}$$

and

$$\frac{A_K}{B_K} = \frac{\left(K\cos(Ka/2) + \frac{\sigma}{a}\sin(Ka/2)\right)}{\left(K\sin(Ka/2) - \frac{\sigma}{a}\cos(Ka/2)\right)}. \tag{A6}$$

As this follows from (A6),

$$\left(\frac{A_K}{B_K}\right)^2 = \frac{\left((Ka)^2 + \sigma^2\right) + \frac{\left((Ka)^2 + \sigma^2\right)^2}{\left[K^2a^2 - \sigma^2\right]}\cos(Ka)}{\left((Ka)^2 + \sigma^2\right) - \frac{\left((Ka)^2 + \sigma^2\right)^2}{\left[K^2a^2 - \sigma^2\right]}\cos(Ka)} \tag{A7}$$

There are two possibilities:

1st. $\cos(Ka) > 0,\ Ka - \sigma > 0$ or $\cos(Ka) < 0,\ Ka - \sigma < 0$ (the same sign), this means $\cos(Ka) = \frac{\left[K^2a^2 - \sigma^2\right]}{\left[K^2a^2 + \sigma^2\right]}$

$$\left(\frac{A_K}{B_K}\right)^2 = \frac{\left((Ka)^2 + \sigma^2\right) + \frac{\left((Ka)^2 + \sigma^2\right)^2}{\left[K^2a^2 - \sigma^2\right]}\frac{\left[K^2a^2 - \sigma^2\right]}{\left[K^2a^2 + \sigma^2\right]}}{\left((Ka)^2 + \sigma^2\right) - \frac{\left((Ka)^2 + \sigma^2\right)^2}{\left[K^2a^2 - \sigma^2\right]}\frac{\left[K^2a^2 - \sigma^2\right]}{\left[K^2a^2 + \sigma^2\right]}} = \infty$$

So, in this case $B_K = 0$

2nd. $\cos(Ka) > 0,$ or $\cos(Ka) < 0,$ (different signs) $\cos(Ka) = -\dfrac{\left[K^2a^2 - \sigma^2\right]}{\left[K^2a^2 + \sigma^2\right]}$
$Ka - \sigma < 0$ $\quad Ka - \sigma > 0$

In this case $\left(\dfrac{A_K}{B_K}\right)^2 = 0$, so, $A_K = 0$

These calculations prove that there are no case, when both coefficients, $A_K$ and $B_K$, are non-zero. So, all the excitations are subdivided into two classes of modes, which profiles are symmetrical (cosinusoidal) or antisymmetrical (sinusoidal) [36]. Evidently, only the modes with symmetrical profiles (case 1 above) are detectable.

We find the wave- vectors along thickness of the film for three observed modes from the experimentally established values $H_{op}{}^n$ and corresponding frequencies $f_n$, using the Kittel formula ((1) in the main text of the article,

$$K_n{}^2 = \left(\dfrac{2\pi}{\gamma} f_n - H_{op}{}^n + 4\pi M\right)\dfrac{M}{2A}, \quad n = 1, 2, 3. \tag{A8}$$

This is easy to see, that when the differences between wave-vectors along thickness of the film are about $\pi/a$, like in our case, the simple approach (presented above) with equal pinning at the both surfaces of the film cannot explain why all three modes are detectable. Really, assume that for the first mode $K_1 a < \pi/2$, i.e., $\cos(K_1 a) > 0$. This mode has a symmetrical profile and can be detectable if $K_1 a > \sigma$. However for the second mode $\cos(K_2 a) < 0$ and $K_2 a > \sigma$, i.e., this mode has sinusoidal form and its intensity is equal to zero, which conclusion does not agree with our experiment. If we assume that $\cos(K_1 a) > 0$, this is easy to prove in the same way, that the third mode is undetectable. This is a typical situation, when there is a reason to assume different pinning at the upper and lower boundaries of the film and to introduce two different pinning constants, $\sigma_1$ and $\sigma_2$. In such a case the boundary conditions can be written in the form (instead of (A1)):

$$\frac{\partial m}{\partial z} - \frac{\sigma_1}{a} m = 0 \Big|_{z=-a/2} \qquad (A9)$$
$$\frac{\partial m}{\partial z} + \frac{\sigma_2}{a} m = 0 \Big|_{z=a/2}$$

And we receive, correspondingly:

$$tg(Ka) = \frac{Ka(\sigma_1 + \sigma_2)}{\left[K^2 a^2 - \sigma_1 \sigma_2\right]} \qquad (A10)$$

$$\cos(Ka)^2 = \frac{\left[K^2 a^2 - \sigma_1 \sigma_2\right]^2}{(Ka)^2 (\sigma_1 + \sigma_2)^2 + \left[K^2 a^2 - \sigma_1 \sigma_2\right]^2} \qquad (A11)$$

We have again two cases:

$\cos(Ka) > 0$

1st. $\quad \left(\frac{A_K}{B_K}\right)^2 = \frac{\left((Ka)^2 + \sigma_1^2\right)\sqrt{1+tg(Ka)^2} + \left((Ka)^2 - \sigma_1^2 + 2(Ka)\sigma_1 tg(Ka)\right)}{\left((Ka)^2 + \sigma_1^2\right)\sqrt{1+tg(Ka)^2} - \left((Ka)^2 - \sigma_1^2 + 2(Ka)\sigma_1 tg(Ka)\right)} \qquad (A12)$

$\cos(Ka) < 0$

2nd. $\quad \left(\frac{A_K}{B_K}\right)^2 = \frac{\left((Ka)^2 + \sigma_1^2\right)\sqrt{1+tg(Ka)^2} - \left((Ka)^2 - \sigma_1^2 + 2(Ka)\sigma_1 tg(Ka)\right)}{\left((Ka)^2 + \sigma_1^2\right)\sqrt{1+tg(Ka)^2} + \left((Ka)^2 - \sigma_1^2 + 2(Ka)\sigma_1 tg(Ka)\right)}. \qquad (A13)$

In the formulae (A12) and (A13) the function $tg(Ka)$ satisfies the relation (A10).

Now both coefficients $A_K$ and $B_K$ in the profiles (A2) can be non-zero and all the corresponding modes can be detectable (of course with different intensity). Using the values, received from (8.A), $K_1 = 0.47\pi/a$, $K_2 = 1.4\pi/a$, $K_3 = 2.3\pi/a$, and the relation (A10), we estimate the set of pinning parameters as $\sigma_1 = 0.1$ and $\sigma_2 = 10$ (or $\sigma_1 = 10$ and $\sigma_2 = 0.1$, as all the formulae are evidently symmetrical). Then, inserting the calculated values $K_n$, $\sigma_1$, $\sigma_2$ (for $n = 1, 3$) into (12A) and $K_2$, $\sigma_1$, $\sigma_2$ into (13A), we find, correspondingly, $(A_1/B_1)^2 = 1.54$, $(A_2/B_2)^2 = 0.4$, $(A_3/B_3)^2 = 3.96$.

**Figures**

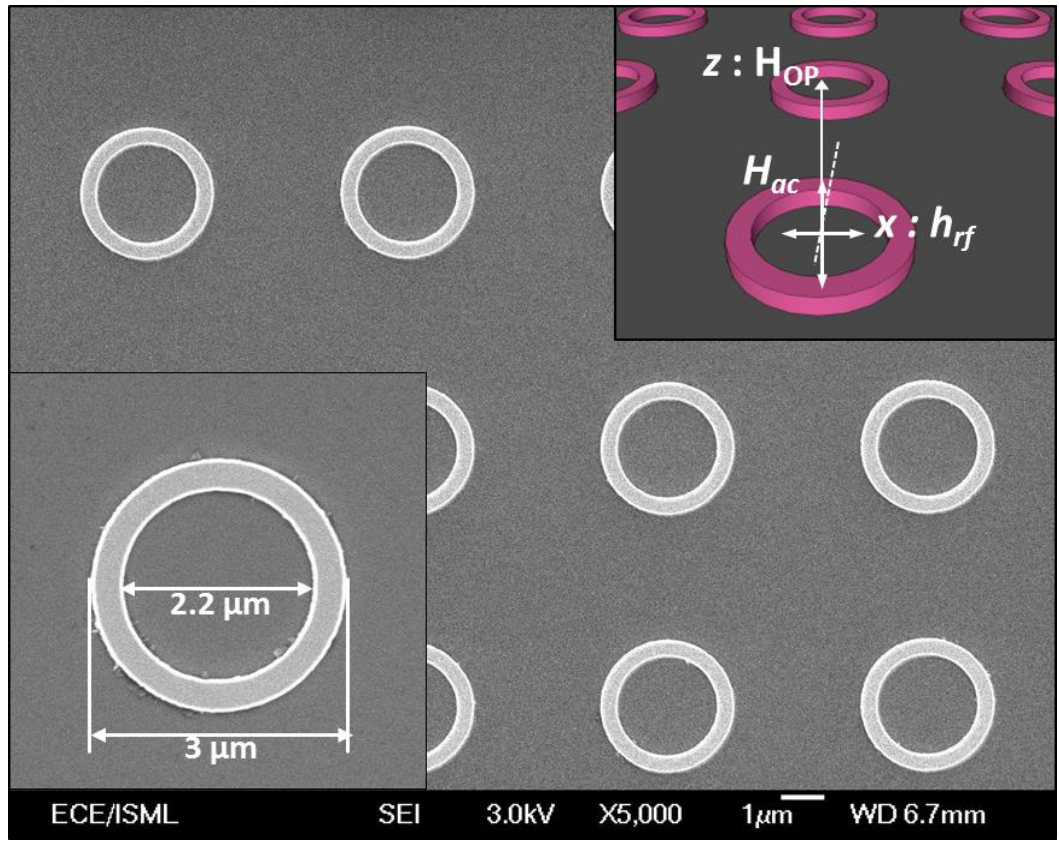

**Figure 1.** SEM image of the periodic array of $Ni_{80}Fe_{20}$ rings. Bottom left inset: SEM image of the isolated ring with indicated dimensions. Top right inset: the geometry of the experiment.

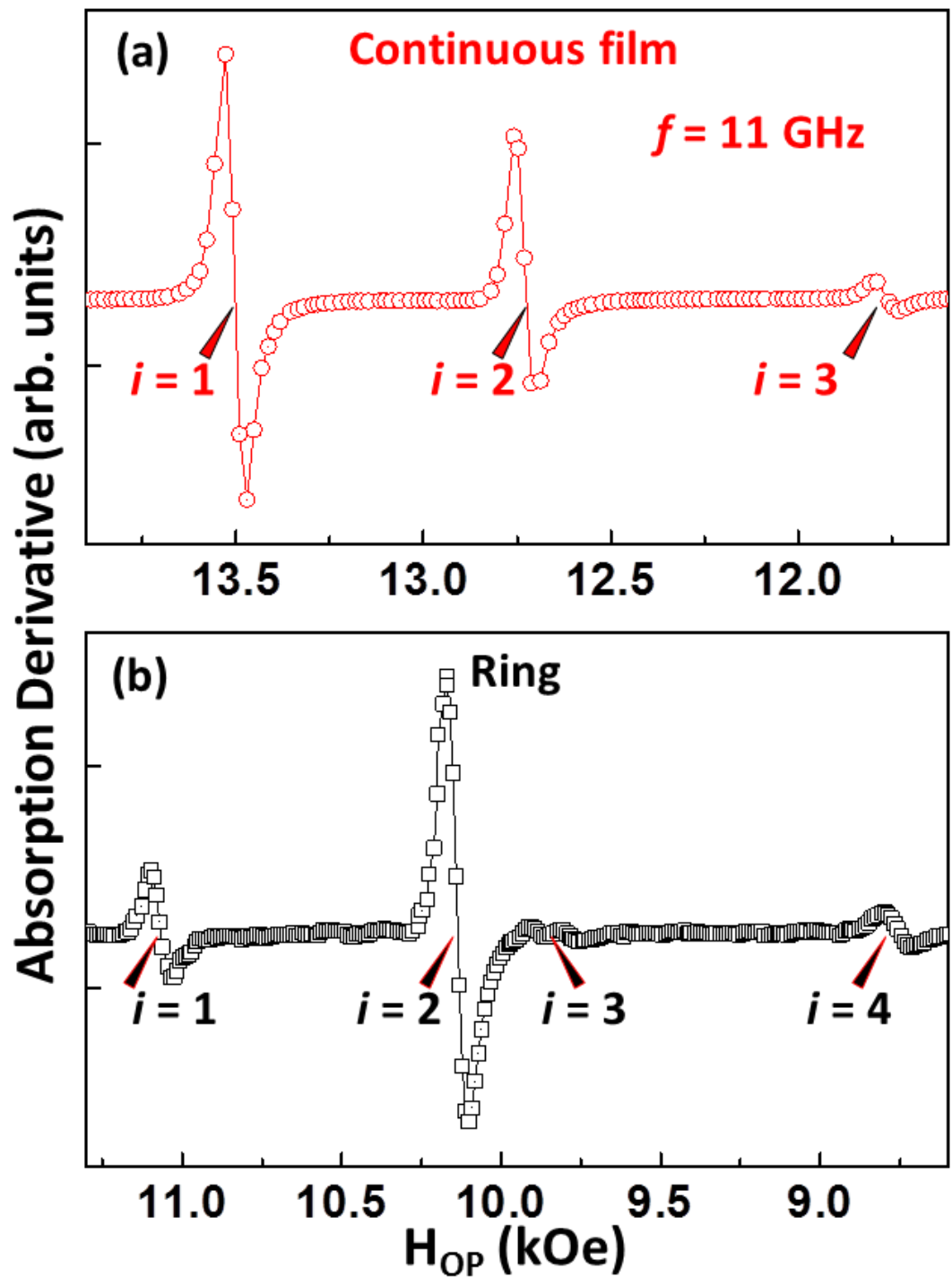

**Figure 2.** The microwave absorption spectra taken at 11 GHz for (a) 100 nm thick continuous film and (b) circular rings.

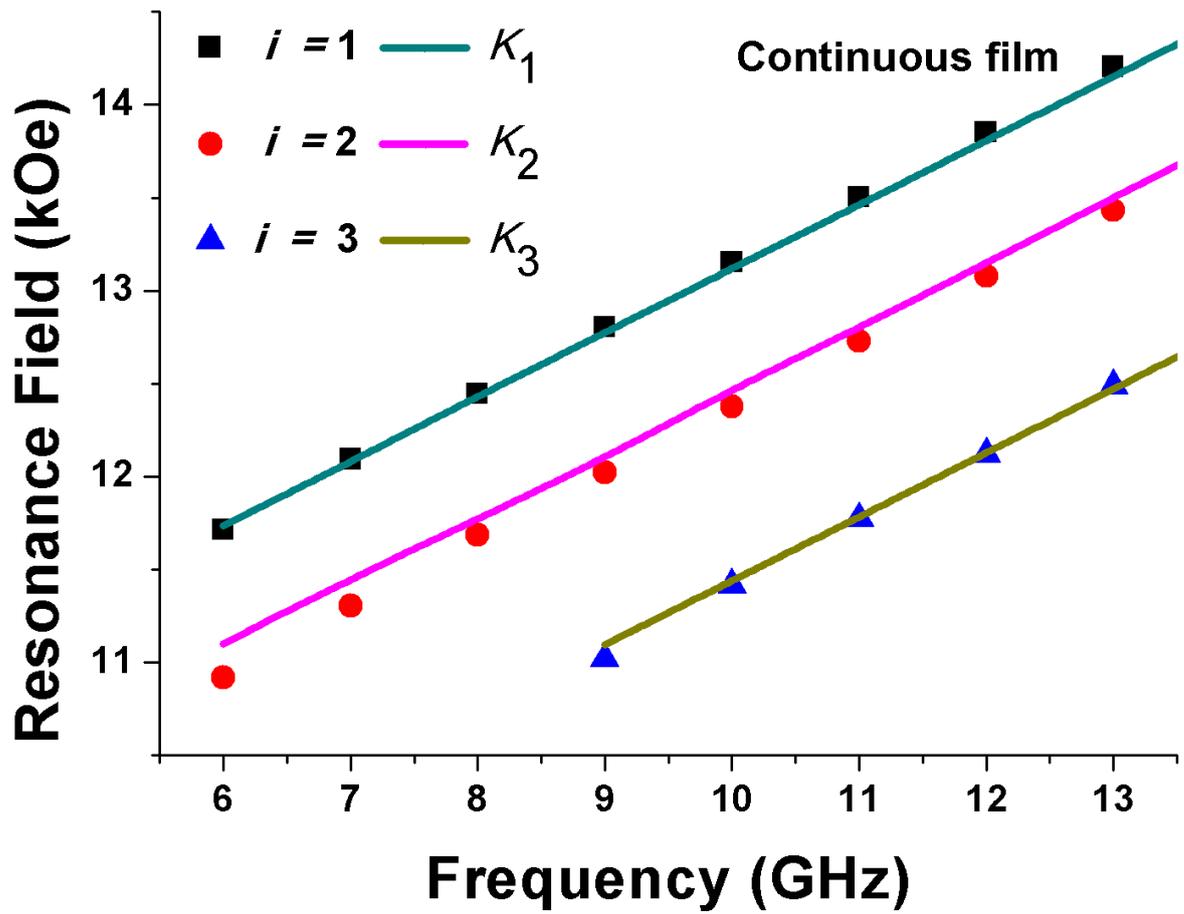

**Figure 3.** Resonance fields extracted as a function of the excitation frequency for the continuous film: squares, dots and triangles - experimental results; lines - calculations by Kittel formula (1) with $K_1 = 0.47\pi/a$, $K_2 = 1.4\pi/a$, $K_3 = 2.3\pi/a$, as explained in the text.

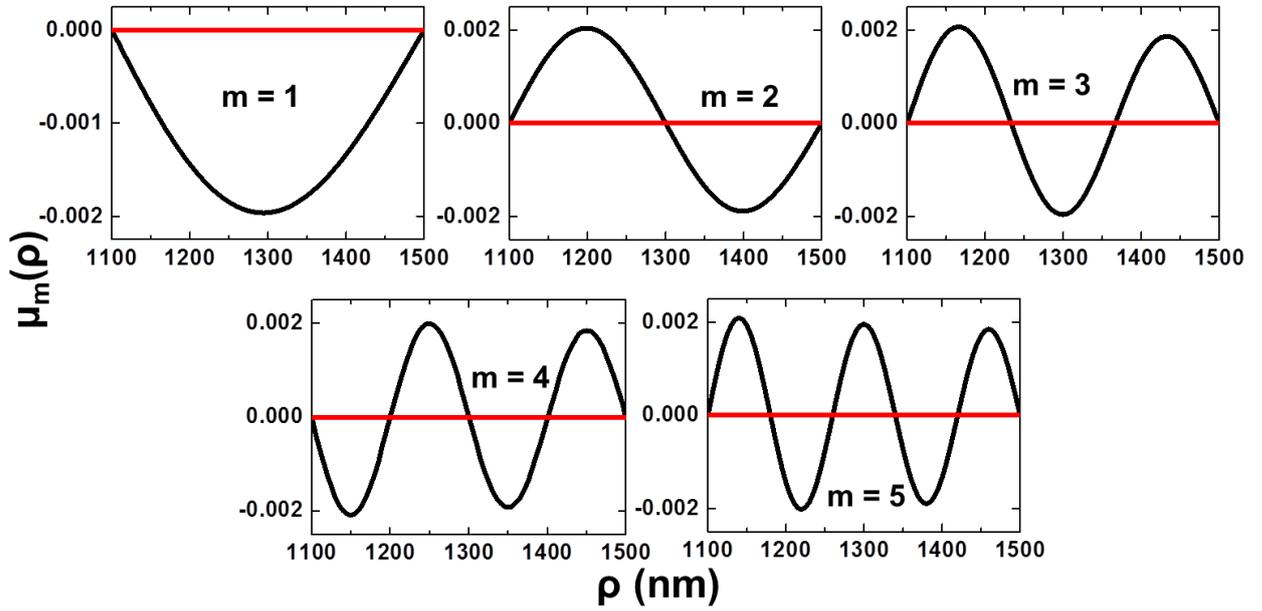

**Figure 4.** Standing spin wave profiles $\mu_m(\rho)$ as functions of the radial coordinate for the first five quantization numbers, calculated by formula (3) with corresponding $C_m = -\dfrac{J_0(\beta_m)}{Y_0(\beta_m)}$ and $\beta_1 = 11.77$, $\beta_2 = 23.554$, $\beta_3 = 35.337$, $\beta_4 = 47.12$, $\beta_5 = 58.902$.

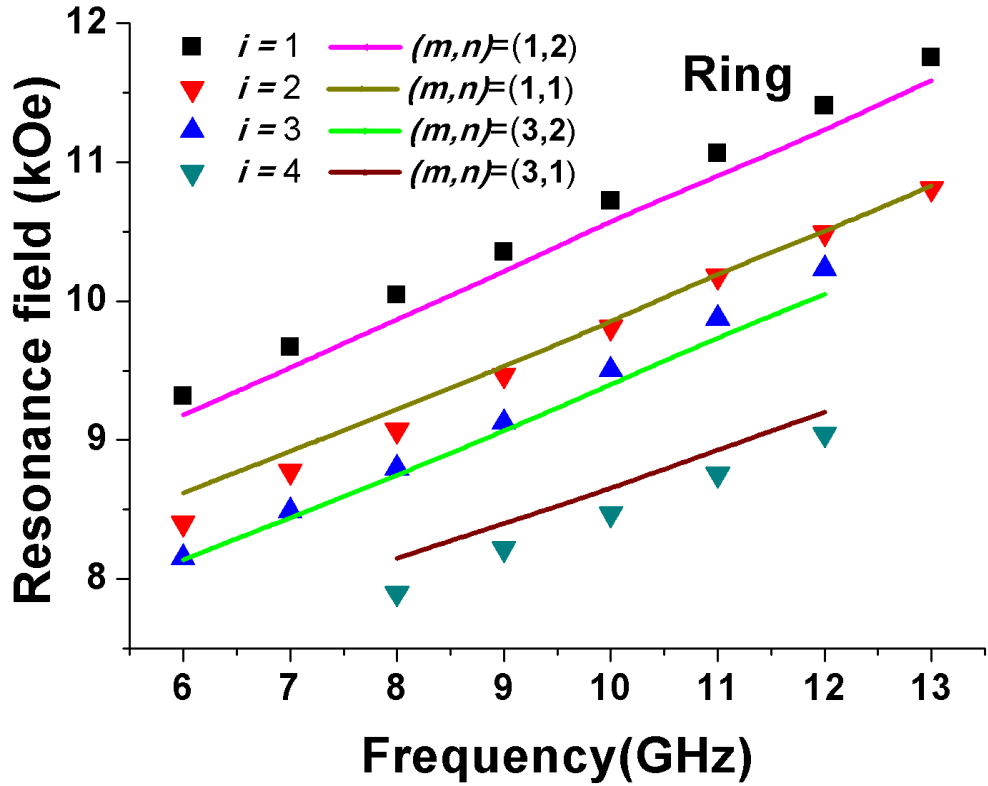

**Figure 5.** Resonance fields extracted as a function of the excitation frequency for the circular ring: triangles and squares - experimental results; solid lines - theoretical calculations by formula (8). The mode with indices $(m, n) = (1, 2)$ is the highest (magenta) line (experimental black squares), the next (second) lower line corresponds to the mode with indices $(m, n) = (1, 1)$ andhas the largest intensity in the FMR experiment (red triangles). The next modes are $(m, n) = (3, 2)$ and $(m, n) = (3, 1)$, as indicated in the Figure.

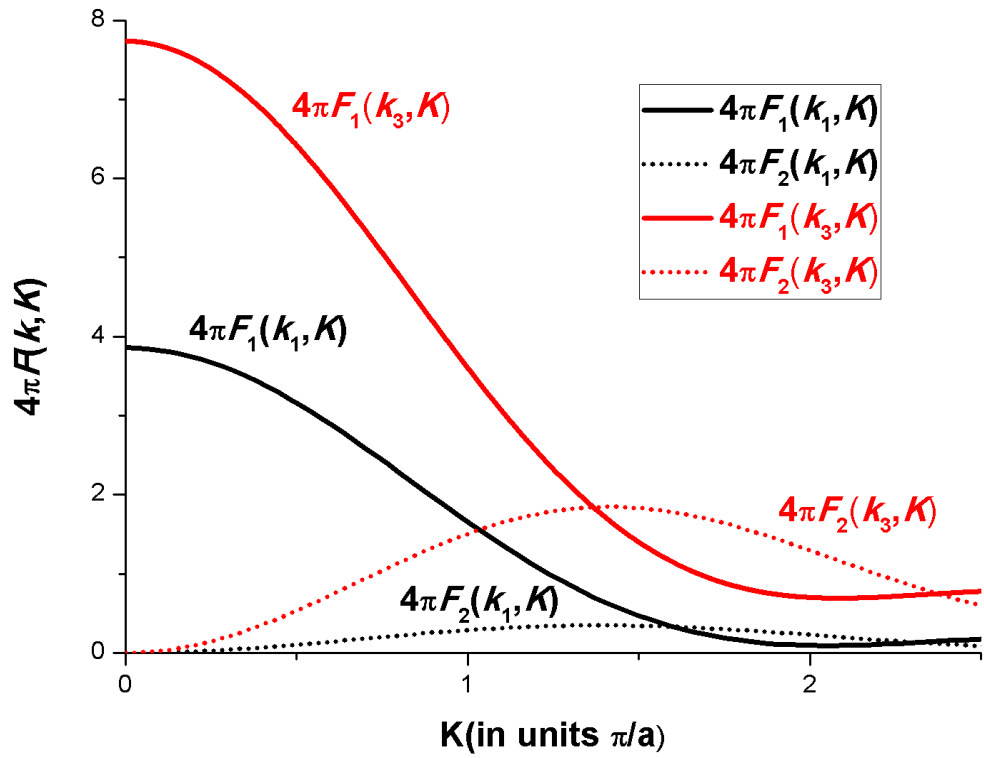

**Figure 6.** Components of the dipole-dipole field $4\pi F_1(k_m, K)$ and $4\pi F_2(k_m, K)$ as a function of out-of-plane wave vector $K$ with fixed in-plane wave vectors $k_1=\beta_1/R$ and $k_3=\beta_3/R$ (calculations by the formulae (10a), (10b)).

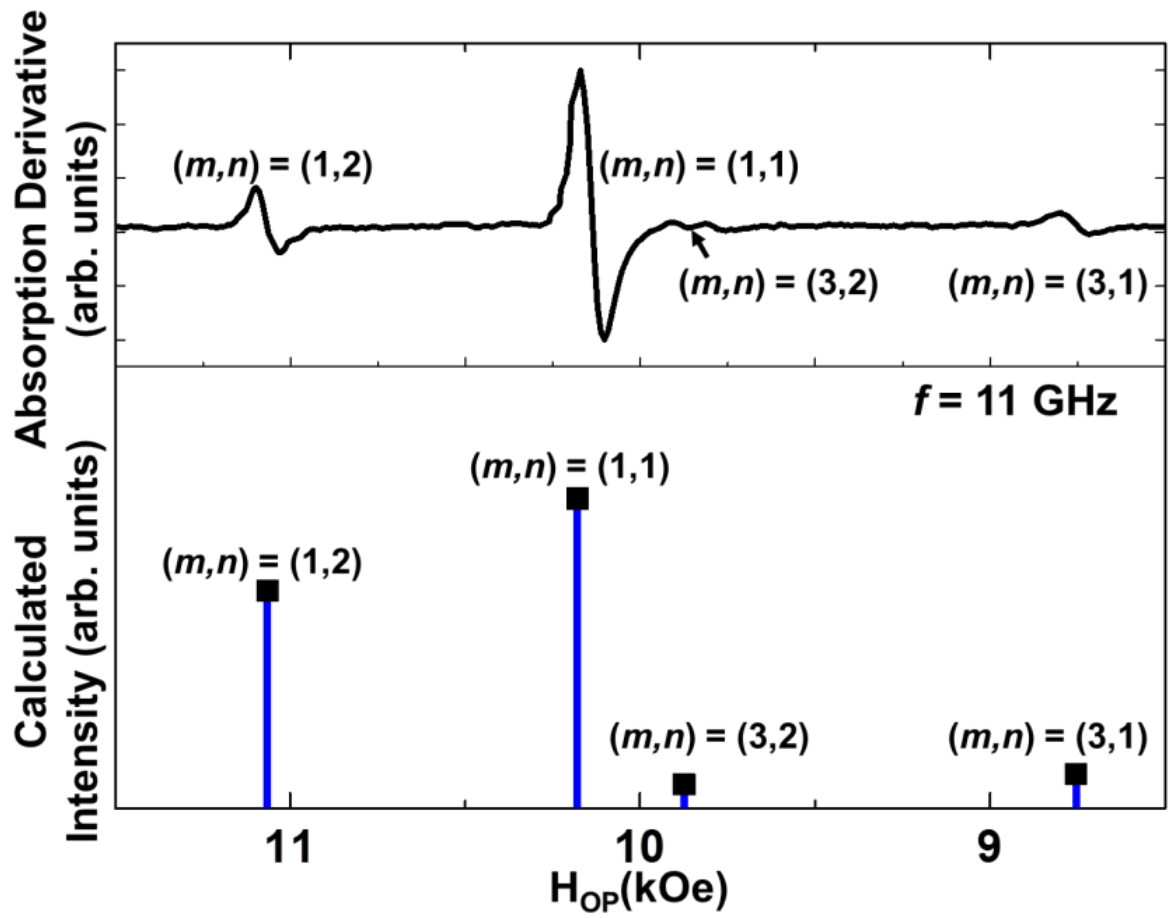

**Figure 7.** Bottom panel: calculated intensity of four spin wave modes observed in rings. Top panel: experimental data for absorption derivative at 11 GHz for comparison.